# An Efficient Communication Protocol for FPGA IP Protection

Farzane Khajuyi[2], Behnam Ghavami[1], Human Nikmehr[2]

[1]Computer Engineering Department, Shahid Bahonar University of Kerman, Kerman, Iran
[2]School of Computer Engineering, Isfahan University, Isfahn, Iran

***Abstract-*** *We introduce a protection-based IP security scheme to protect soft and firm IP cores which are used on FPGA devices. The scheme is based on Finite State Machin (FSM) obfuscation and exploits Physical Unclonable Function (PUF) for FPGA unique identification (ID) generation which help pay-per-device licensing. We introduce a communication protocol to protect the rights of parties in this market. On standard benchmark circuits, the experimental results show that our scheme is secure, attack-resilient and can be implemented with low area, power and delay overheads.*

## 1. Introduction

The number of published journal papers concerning FPGA implementation of various applications has increased [1]. The extensive use of FPGAs for hardware realization of almost any digital design in the industry is mainly due to the significant progress in developing new FPGA technologies [2].

In order to reduce cost and time to market, and to develop more optimized designs todays system designers are forced to exploit more Intellectual Property (IP) cores, which are designed and sold as HDL codes (soft IP), optimized synthesizable RTL (firm IP) or GDSII (hard IP) by IP vendors [3].

Even though IP cores are sold through either pay-per-device licensing or up-front license fees, which allows system designers to use IP core in any FPGA devices, IP vendors prefer the former approach, which restricts the user to utilize only licensed FPGA devices [4].

IP security schemes should enforce pay-per-use licensing as well as protecting IP from piracy attacks. These schemes are categorized by VSI Alliance as [5]:

- *Deterrent-based approaches*, where IP vendors can only deter adversaries from misuse and piracy but prevent. Some approaches in this category include: patents, copyrights, trade secrets, contracts and lawsuits.
- *Protection-based approaches*, where the IP providers can take actions to prevent piracy and unauthorized use of their IPs. Licensing agreements and encryption are examples of this class of IP protection methods.
- *Detection-based approaches*, where IP vendors can detect unauthorized use of their properties. Examples of the detection-based approaches include foundry IP tracking or tagging, digital signatures (such as digital fingerprinting and digital watermarking) and noise fingerprinting.

In this paper we propose a protection-based method which locks IP by introducing additional states to the original IP's FSM and enforces the pay-per-use licensing by using the Physical

Unclonable Function (PUF) responds for IP unlocking. PUFs are challenge-response systems based on process variation [6]. These functions get a set of input challenges and map a set of responses to them. Although the responses are measurable, they are not predictable due to the fact that they are generated based on the process variation and it is why PUFs cannot be copied or cloned. A wide range of such functions have been proposed in the literature [7,8] from which FPGA vendors can pick one that suits their device's characteristics and implementation constraints. However, the proposed method is specification independent and can be used by various FPGAs and PUFs.

The rest of the paper is organized as follows. We briefly survey the related works in Section 2 followed by introducing the proposed approach and its procedure in Section 3. Section 4 reports the experimental results, analysis and comparison results. Finally, a thorough conclusion is given in Section 5.

## 2. Related Work

Configuration content encryption, one of the primary protection-based schemes to provide IP security, was invented by Austin in 1995 [9]. Since then, a variety of methods has been proposed using IP encryption. All these approaches are based on the following three steps:

1. IP core encryption
2. Selling the IP as a black box
3. Decryption while loading on FPGA.

While the main difference between these methods concerns how they generate and maintain the encryption key, their main drawback is the key management itself [10]. Having recalled the license agreement as an IP protection scheme, a common approach for implementation of this scheme is through IP obfuscation, which is performed by making changes in some parts of the IP's code. The obfuscation methods protect IP core from piracy attacks as well as enforcing authorized use. These alterations can take place in the control (FSM), data-path, memory and/or I/O units so that it does not affect the original IP's function. These methods usually propose communicational protocols for involved parties to make the key and IP exchanging process secure.

The idea of FSM obfuscation is first introduced and implemented by Alkabani and Koushanfar to protect hard IPs [11] [12]. They propose a method in which the FSM goes to its normal function just after applying a specific input sequence (called the "key"). In this approach some states of the FSM are selected randomly or heuristically and repeated in certain number of times. Even the evaluation results show that power, delay and area overheads are improved when the repeating states are chosen heuristically (compared to the random counterparts) the average power overhead of up to 84%makes Alkabani and Koushanfar's heuristic-based approach practically infeasible. As the first idea aims to protect hard IPs, their proposed communicational protocol includes parties in both hard IP and ASIC markets (even though the protocol can be extended to support soft IP market as well, practically it is not possible due to its complexity). Alkabani and Koushanfar in [13] introduce an optimized version of their previous method that is capable of protecting multiple IPs. However, power and area overheads are still too high to make it practical.

Another method proposed by Chakraborty and Bhunia [14] employs obfuscation by modifying the circuit's gate-level netlist. Although area overhead is improved by this method, it is just applicable to the gate-level netlist IPs and includes an internal nodes choosing process. Again, this technique is basically developed to be used on hard IPs and its proposed communicational protocol is not suitable for soft IPs.

The first FSM obfuscation method developed to protect soft IPs in FPGA-based designs is proposed in 2015 [15]. This approach adds an FSM to the original FSM and binds it to the PUF which is integrated into the FPGA device. The added FSM contains M layers of new states with m states in every even-numbered and one state in every odd-numbered layers. The single state in an odd-numbered layer connects to all m states in the next (even-numbered) layer while each of the m states in any even-numbered layer connects to the single state in next (even-numbered) layer. Every odd transitional step (from an odd-numbered to the next even-number layer) is encoded using $\log_2 m$ bits of the PUF's output. However, every even transitional step is represented as a $(\log_2 m)$-bit binary number obtained by do XOR operation $\log_2 m$ bits of the PUF's output and $\log_2 m$ bits of the license. Now, once the IP core is used on an authorized FPGA device, the added FSM is unlocked by the PUF and license after M transitions. Fig.1 illustrates the unlocking mechanism proposed in this method and how the FSM is added using an example where the added FSM contains two layers ($M = 2$) with four states in the second layer ($m = 4$).

Despite its advantages, this method is not practically feasible duo to its high area overhead [16], failure probability to preparing security and complexity, and communicational protocol cost.

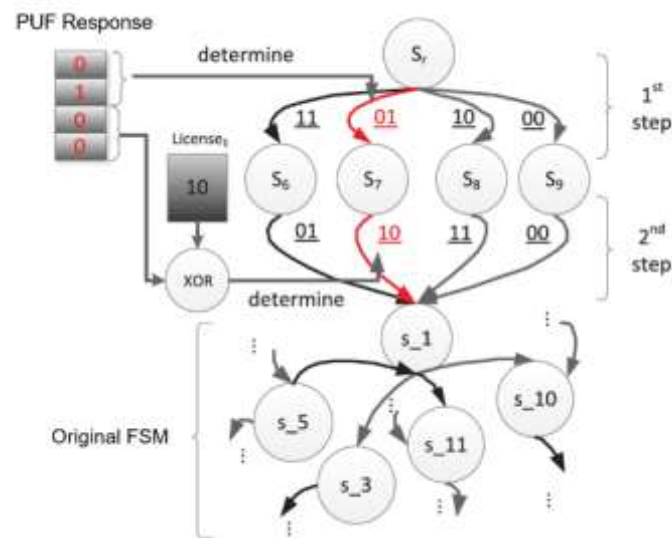

Fig. 1. Unlocking mechanism proposed in [15].

Investigations reveal that almost all previous FSM obfuscation methods reported in the literature suffer from two main drawbacks; high area and power overheads as well as complex communicational protocol. In this paper we propose a new FSM obfuscation approach to address these issues.

## 3. Proposed Method

The proposed obfuscation scheme binds IP core to an authorized FPGA by modifying its FSM and creating a Boosted FSM (BFSM). BFSM is created by adding dummy states and transitions to original FSM. Dummy transitions are dependent on PUF responses. So the IP core won't have its correct functionality unless it is used in an authorized FPGA. The proposed dummy FSM contains a black hole FSM. Black hole FSM is a FSM with the black hole states. Exit from these states is impossible, regardless of the input sequence. Adding these states to BFSM make random guessing attacks impossible [11]. We show black hole states in gray color in this paper. The proposed IP protection scheme works in the following way:

### 3.1. *Communication protocol*

The communication protocol is designed to protect IP vendor as well as being simple and practical. Figure 2 shows this protocol. It can be described as follows.

1. FPGA vendor produces FPGAs. Each FPGA contain a PUF and has a specific ID. IP vendor also produces IPs with IDs.
2. FPGA vendor sends challenge-response (C-R) pairs of each FPGA to IP vendor. Challenges are stored in a non-volatile memory and will be configured on FPGA as it powers on.
3. FPGA vendor sells her/his product to system designer.
4. System designer send IDs of both her/his required IP and FPGA to IP vendor.
5. IP vendor finds C-R pairs of FPGA and obfuscates IP's FSM.
6. IP vendor sells the locked IP and license to system designer.
7. System designer use license to unlock the IP.

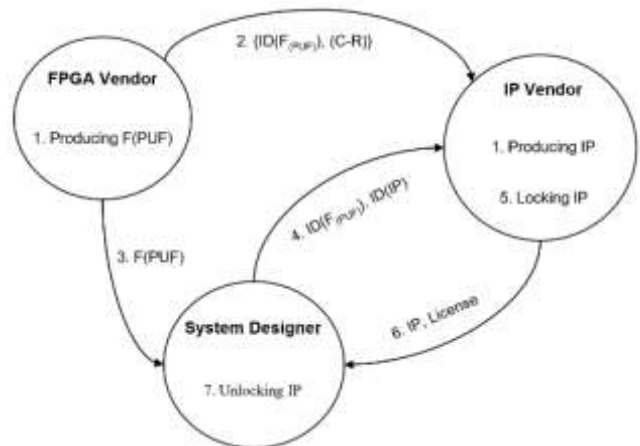

Fig.2. proposed communication protocol

### 3.2. *Modification of FSM*

To modify the FSM, as mentioned before, we insert a dummy FSM in the IP. To create dummy FSM, we add n states and h black hole states to the original FSM. Each of n state is connected to all h black hole states. But there is only one transition from each state to one another normal state. So there are $h + 1$ transitions from each state. The number of bits that determine transitions can be computed by:

$$b = \log_2(h + 1) \tag{1}$$

Transitions from even-number states ($S_{n0}$, $S_{n2}$, …) are determined by $b$ bits of PUF response. Transitions from odd-number states ($S_{n1}$, $S_{n3}$, …) determiner is the result of XOR $b$ bits of PUF response and $b$ bits of license. For instance, transition determiner for $S_{ni}$ is $(r_{i+1}r_i)$ when $i$ is even and $(r_i r_{i-1} xor l_i l_{i-1})$ when $i$ is odd. Where $r_i$ and $l_i$ shows the $i$th bit of PUF response and license respectively and $b = 2$.

Figure 3 illustrates a BFSM and its transition determiner mechanism. The original FSM consists 7 states $S_0$ to $S_6$. We insert dummy FSM with 4 normal states $S_{n0}$ to $S_{n3}$ and 3 black hole states $h_0$ to $h_2$. Although black hole states can connect together in any way, there is no way out from black hole FSM. This FSM is shown in gray color.

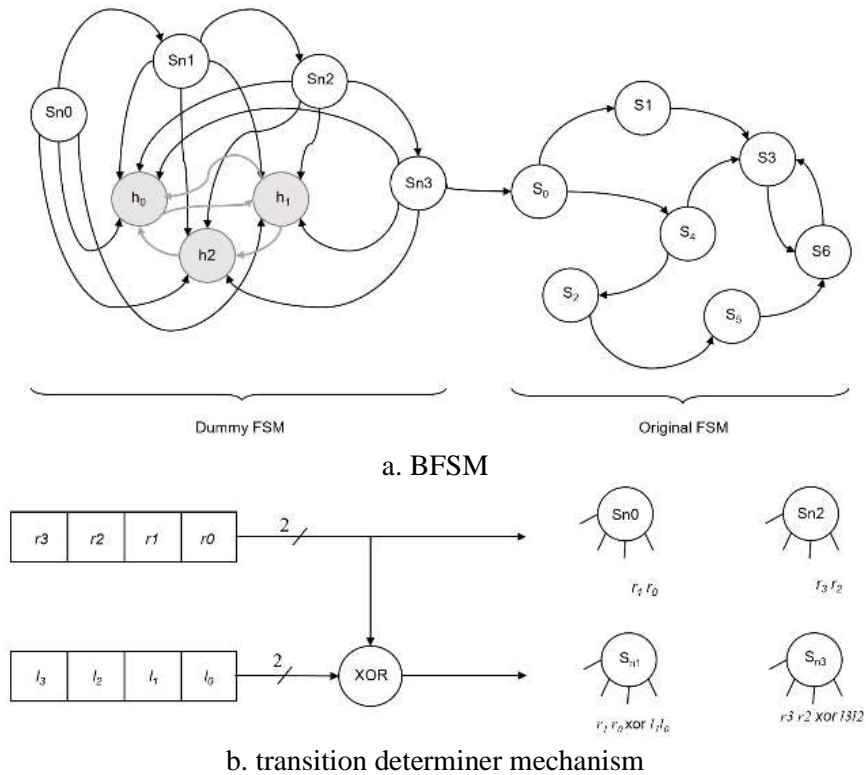

a. BFSM

b. transition determiner mechanism

Fig.3. Modification example for an FSM with 7 states

### 3.3. *Determination of the dummy state's number*

In this subsection, we aim to compute minimum number of states (*SN*) for a given license. As mentioned before, every *b* bits of license are used to specify two transition steps. There are *n* transition steps in a dummy FSM with *n* states. So the length of license bits (*L*) can be formulated as fallow.

$$L = \frac{nb}{2} \qquad (2)$$

Equation (2) can be rewritten as Eq. (3) which clearly demonstrates that *n* is even. And so the PUF response and license are equal in length since every transition is determined by using *b* bits of PUF response and license alternatively.

$$n = \frac{2L}{b} \qquad (3)$$

Each dummy FSM consists of $n + h$ states. By using equations (1) and (3) the states number can be rewritten as follow.

$$SN = (2^b - 1) + \left(\frac{2L}{b}\right) \quad (4)$$

We can calculate *b* that leads to minimum *SN* as bellow.

$$\frac{dSN}{db} = 0 \rightarrow$$
$$b + 2\log_2 b = \log_2 \frac{2L}{\ln 2} \rightarrow$$
$$2^b \times b^2 = \frac{2L}{\ln 2} \quad (5)$$

Now, the number of **dummy** states can be computed according equations (1) and (3).

### 3.4. *An example*

Suppose Mina is a system developer. And she has bought a FPGA which its ID is "123". Now, she needs a special IP core. She seeks her required IP core from all IP cores which produced by IP vendor. And sends its ID in addition to "123" to IP vendor. IP vendor, who received IDs, search her/his database for C-R pairs of "123" FPGA. (According to steps 1 and 2 of communication protocol, FPGA vendor had sent FPGAs' ID and C-R pairs to IP vendor). For example, consider PUF response is "001011". So IP vendor can determine *b*, *n* and *h* according to equations (5), (3) and (1) as 2, 6 and 3 respectively. Then, IP vendor lock the IP by adding dummy FSM to original FSM, synthesizes the IP and sends locked IP core (configuration bitstream) and license to Mina, who can use this IP just in FPGA "123". Figure 4 and table 1 illustrate dummy FSM and parameters value for this example. $S_{n0}$ and $S_0$ are the first states of dummy FSM and original FSM, respectively. BFSM will be unlocked after passing sequence of states $S_{n0}$, $S_{n1}$, $S_{n2}$, $S_{n3}$, $S_{n4}$, $S_{n5}$.

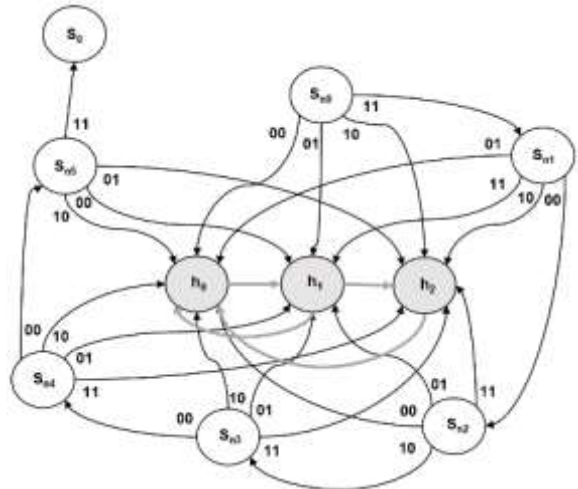

Fig.4. Example dummy FSM

Table1. Example parameters value

| parameter | value |
|---|---|
| L | 6 |
| b | 2 |
| n | 6 |
| h | 3 |
| PUF response | 001011 |
| license | 111011 |

# 4. Experimental Results

In this section, at first, we present simulation results of the proposed scheme and compare them with the results reported in [15] as the first obfuscation-based scheme for IP protection in FPGA-based systems. Then, we show why the scheme presented in [15] is not completely secure. At last, we mention some attacks and describe how our scheme is secure against them.

*4.1. Delay, power and area overheads*

The proposed scheme is implemented on virtex5 xc5vlx20t FPGA using ISE 14.7 for synthesis. Experiments were conducted on a 3.6 GHz intel corei7 processor machine with 8GB RAM. Added FSMs were written in Verilog. We used sequential benchmarks from the MCNC'91 set and turn them from kiss2 format to Verilog by kiss2vl tool [17].

Considering two different license length, added FSMs for ten benchmark circuits are simulated according to our scheme as well as proposed scheme in [15]. Table 2 illustrates the number of added states in both schemes when license length is four and six. "L" are used to show license length.

Table 2. state number in tow schemes

| L | Our scheme | Ref.[15] |
|---|---|---|
| **4** | 7 | 10 |
| **6** | 9 | 15 |

We use ten sequential circuits of MCNC'91 benchmark as IP cores which should be protected. Considering license length 4 and 6, we add dummy FSMs of each scheme to IPs and synthesize them. We use xPower toll to evaluate the power consumption. Tables 3 and 4 show the results.

Table 3. Synthesize results where $L = 4$

| Circuits | Our scheme | | | | Ref. [15] | | | |
|---|---|---|---|---|---|---|---|---|
| | #LUTs | #Slices | Delay (ns) | Power (mw) | LUTs# | #Slices | Delay (ns) | Power (mw) |
| *dk16* | 39 | 13 | 1.985 | 376 | 45 | 15 | 1.830 | 560.49 |
| *S832* | 90 | 31 | 2.108 | 406 | 135 | 70 | 3.224 | 561.21 |
| *S510* | 48 | 20 | 2.057 | 376 | 64 | 23 | 2.184 | 560.49 |
| *S820* | 100 | 45 | 2.034 | 401 | 129 | 55 | 2.639 | 560.73 |
| *styr* | 164 | 67 | 2.481 | 364 | 200 | 65 | 2.38 | 560.49 |
| *S1488* | 157 | 63 | 2.260 | 440 | 193 | 77 | 2.261 | 560.97 |
| *S1494* | 162 | 68 | 2.512 | 415 | 197 | 78 | 2.225 | 560.49 |
| *planet* | 145 | 60 | 2.227 | 432 | 182 | 75 | 2.383 | 560.73 |
| *S298* | 302 | 136 | 3.037 | 401 | 406 | 148 | 4.088 | 560.49 |

Table 4. Synthesize results where $L = 6$

| Circuits | Our scheme | | | | Ref. [15] | | | |
|---|---|---|---|---|---|---|---|---|
| | #LUTs | #Slices | Delay (ns) | Power (mw) | #LUTs | #Slices | Delay (ns) | Power (mw) |
| *dk16* | 46 | 15 | 2.090 | 335 | 54 | 17 | 2.119 | 560.49 |
| *S832* | 137 | 64 | 2.630 | 376 | 137 | 60 | 2.838 | 560.49 |
| *S510* | 55 | 18 | 2.003 | 379 | 75 | 27 | 2.125 | 560.49 |
| *S820* | 111 | 51 | 2.526 | 355 | 137 | 60 | 2.84 | 560.49 |
| *styr* | 172 | 66 | 2.557 | 379 | 203 | 58 | 2.819 | 560.73 |
| *S1488* | 153 | 57 | 2.070 | 448 | 199 | 84 | 2.482 | 560.73 |
| *S1494* | 178 | 78 | 2.503 | 443 | 198 | 78 | 2.47 | 560.49 |
| *planet* | 159 | 70 | 2.415 | 452 | 176 | 78 | 2.631 | 560.73 |
| *S298* | 302 | 112 | 3.010 | 374 | 256 | 84 | 3.214 | 560.73 |

Fig. 5 and 6 demonstrate the average delay, power and area overheads for our scheme compared to the scheme that proposed in [15] when license length is 4 and 6 respectively. It is interesting to note that in many cases the power overhead is negative. When $L = 4$ the average delay overhead is 4.49% and 17.77% for our scheme and [15] respectively, the power overhead is -3.48% and 0.03%, and area overhead ( occupied LUTs ) is 19.95% and 52.02%. The results show that our scheme outperforms the scheme proposed in [15] on average.

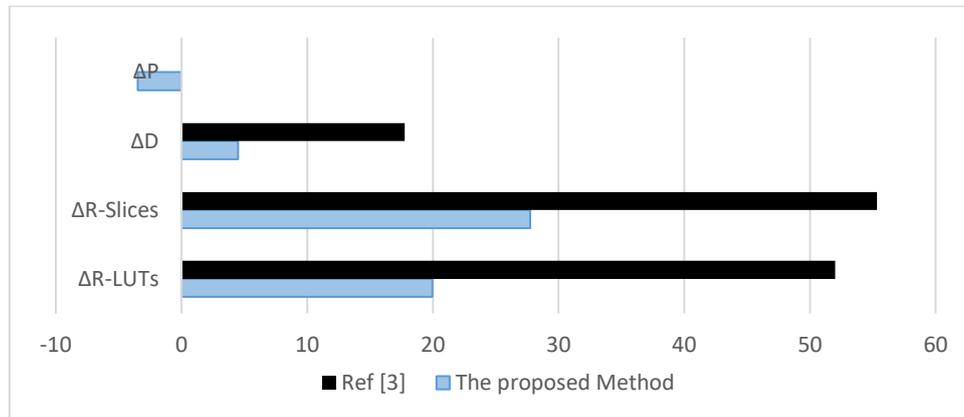

Fig.5. Average overhead of both schemes where license length is 4 bits

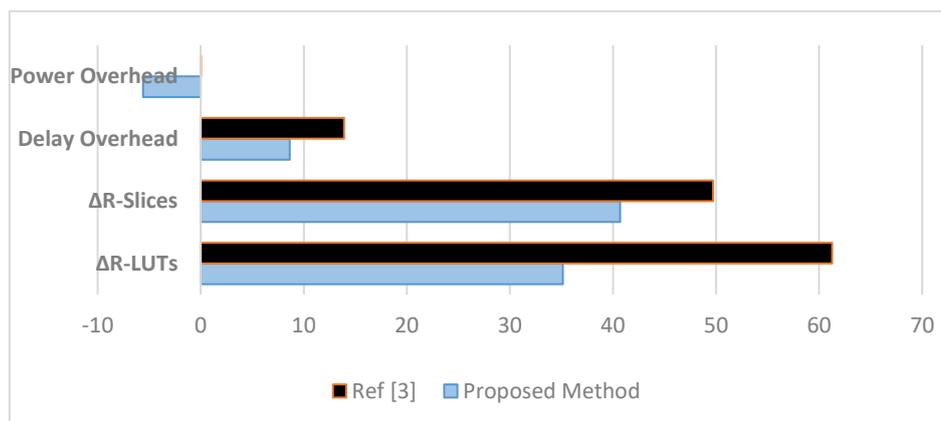

Fig.6. Average overhead of both schemes where license length is 6 bits

*4.2. Overhead of the high security solution*

The license has to be at least 128-bit long to make the scheme secure against brute force attacks [16]. Increment of license length increases the number of added states and area overhead. To compare the area overhead of two schemes, we calculate the number of added states when both schemes are secure against brute force attacks($L = 128$). Considering $L = 128$, the parameters "m" and "M" are equal to 3 and 178 respectively [16]. And so added states number is 365 in ref [15] scheme. In our scheme, we can compute b, n and h equal to 4, 64 and 15 from equations 5,3 and 1 respectively. And so added states number is $64 + 15 = 79$ in our scheme. The more dummy states a scheme adds, the more overhead the scheme has. Therefore, in high security level, our scheme outperforms the scheme proposed in ref [15] by cutting the added states number into quarter.

*4.3. Probability of unlocking IP core on unauthorized FPGA device when license is disclosure*

Even if an adversary found the license, he/she would not be able to use the IP core on any FPGA device except the authorized one which contain correct PUF responses. For instance, consider our example in former section. Suppose Mina, who has the License, wants to reuse the IP core on another FPGA device without IP vendor permission. what is the probability of her success? To answer this question, we simulate both ref [15] and our scheme. We use ModelSim PE student edition 10.4a for simulation and write codes in VHDL. Figures 7 and 8 show result for ref [15] and our scheme respectively.

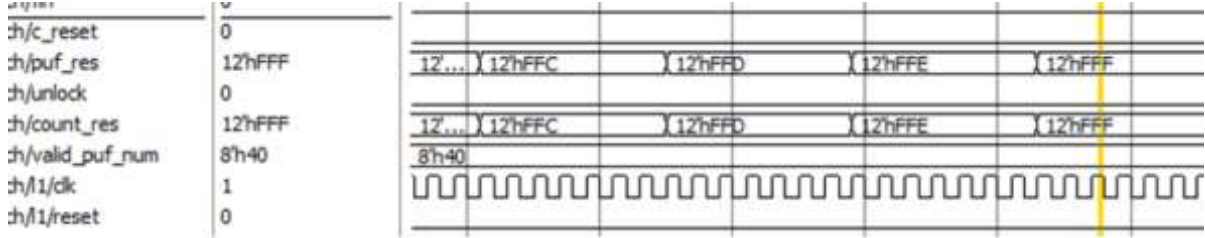

Fig. 7. Simulation result for valid PUF responses in ref [15] scheme

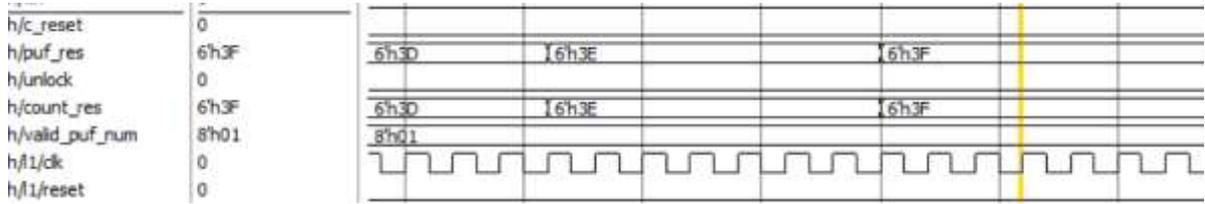

Fig. 8. Simulation result for valid PUF responses in our scheme

To examine all PUF responses, we use 6-bit and 12-bit binary counter for our scheme and ref [15] scheme respectively. As we reduce the required PUF response length to half, length of license and PUF response is equal in our scheme while ref [15] scheme should use PUF responses with double length for the same license. "puf_res" signal is connected to "count_res" signal which is the output of binary counter and get all possible responses. "valid_puf_num" signal illustrates the sum of PUF responses which can unlock the IP. (while license is a constant value) as figures show, after applying all possible PUF responses, that means $(FFF)_{16}$ and $(3F)_{16}$ for "puf_res" signal in Fig. 7 and 8, value of this signal is equal to $(40)_{16}$ and 1 for ref [15] and our scheme respectively.

The number of valid PUF response is equal to the number of existing paths from first state of dummy FSM to its last state (unlock state). There is only one path from $S_{n0}$ to $S_{n-1}$ in our scheme. In ref [15] scheme, as Fig. 1 shows, to make a path from first state to the end, we have $m$ option in each odd-numbered layers and only one option in every even-numbered layers. Every dummy FSM in this scheme has $M$ layers which half of them are odd-numbered. So we have $m^{\frac{M}{2}}$ different paths from first state to unlock state. As PUF response length is $M[\log_2 m]$ bits [15], sample space is equal to $2^{M[\log_2 m]}$. So we can simply compute the probability of unlocking the IP core by an unauthorized FPGA device (when license is known) as Eq. (6):

$$P = \frac{m^{\frac{M}{2}} - 1}{2^{M[\log_2 m]}} \qquad (6)$$

As there is only one path from first state of dummy FSM to its last state, the equation (6) is rewritten as Eq. (7) for our scheme.

$$P = \frac{1-1}{2^{M[\log_2 m]}} = 0 \tag{7}$$

Therefore, the probability of unlocking the IP core by an unauthorized FPGA device is zero in our scheme.

*4.4. Attacks*

Here, we discuss some of the possible attacks on our scheme and the resistance of it against them.

- Brute-force attack: The adversary tries to guess the correct license by testing random licenses. But it is not a feasible attack. As sample space is an exponentially function of license length, by increasing the length of license probability of correct guessing will be almost zero. The advantage of proposed scheme, compared to ref [15] scheme, is lower area overhead while increasing license length.
- Reverse engineering of added FSM: this attack is almost impossible for the proposed method and previous methods because removing added FSM from original FSM is a very complicated computing problem [15,12,13]. Besides that, since our scheme contains black holes, it is more resistant against this attack [15,12].
- Simulating PUF: system developer may simulate PUF behavior after buying IP core and license ones and use it for next times on unauthorized devices. But this attack is impossible duo to modern technology of fabrication [13,14].
- Tapping PUF responses: proposed method is resist against this attack since PUF responses are temporary.
- Side channel attack: these attacks get information about devices by statistical analysis of time, power consumption or electromagnetic emanation. Protecting methods based on PUF usually are not resistant against these attacks [18].

## 5. Conclusion

We present an IP protection scheme that binds IP core to specific FPGA device by utilizing PUF and modifying the FSM of circuit. The scheme also contains a protocol for communication of involved parties (FPGA vendor, IP vendor and system designer) and helps to preserve the rights of all of them. Experimental evaluations on sequential circuits of MCNC'91 benchmark set demonstrate that this scheme is capable of providing high levels of security at low overhead and cost and being utilized in industrial-strength designs.